\documentclass[showpacs,aps,amssymb,floatfix,prd,amsmath,preprintnumbers]{revtex4}
\usepackage{epstopdf}
\usepackage{capt-of} 
\usepackage{hyperref}
\usepackage{graphicx}  
\usepackage{dcolumn}   
\usepackage{titlesec} 
\usepackage{bm}
\usepackage{setspace}
\begin{document}
\input epsf.tex	

\title{\bf LRS Bianchi type-I bulk viscous cosmological models in $f(R,T)$ gravity}

\author{
Parbati
Sahoo\footnote{Department of Mathematics, Birla Institute of
Technology and Science-Pilani, Hyderabad Campus, Hyderabad-500078,
India,  Email:  sahooparbati1990@gmail.com.}, Raghavender Reddy
\footnote{Department of Mathematics, Birla Institute of
Technology and Science-Pilani, Hyderabad Campus, Hyderabad-500078,
India,  Email: vedire8228@gmail.com.}}

\affiliation{ }

\begin{abstract}
\begin{center}
\end{center}

We have studied the locally rotationally symmetric (LRS) Bianchi type-I cosmological model in $f(R,T)$ gravity ($R$ is the Ricci scalar and $T$ is the trace of the stress energy tensor) with Bulk viscous fluid as matter content. The model is constructed for the linear form $f(R,T)=R+2f(T)$. The exact solution of field equations is obtained by using a time varying deceleration parameter $q$ for a suitable choice of the function $f(T)$. In this work, the bulk viscous pressure $\bar{p}$ is found to be negative and energy density $\rho$ is found to be positive. The obtained model is anisotropic, accelerating and compatible with the results of astronomical observations. Also, some important features of physical parameters of this model have been discussed.
\end{abstract}



\maketitle


\section{Introduction}
The current understanding of the universe is that it is undergoing a phase of expansion. Based on astronomical observations and results, it is estimated that the rate of expansion is increasing with time. A few years ago, the high redshift super-novae experiments \cite{Garnv98, Riess98, Perl99, Perl97, Bennet2003} and cosmic microwave background radiation \cite{Spergel03, Spergel07} have provided the necessary evidence for the accelerated expansion of the universe. It is believed that within the framework of the standard cosmological model, a most common reason for this expansion is due to an unknown form of energy, called ``dark energy" (DE), which is an exotic matter with negative pressure. However, knowing the real nature of dark energy has always been a challenge to scientists. According to different estimations, it occupies larger share (73\%) of total energy of the universe, while dark matter constitutes 23\%, and usual baryonic matter occupy about 4\%. The simplest candidate for dark energy is the cosmological constant ($\Lambda$). Many theories have established a very lucid relationship between dark energy and expansion of the universe. In order to explain this accelerated expansion of the universe two different ways are usually chosen. One is by constructing various dark energy candidates like cosmological constant \cite{padmnnavan03}, quintessence \cite{farooq11,martin08}, phantom energy \cite{nojiri03, jamil11}, k-essence \cite{chilba00}, tachyon \cite{Padmanavan02}, Chaplying gas \cite{Bento02} and cosmological nuclear energy \cite{Gupta10}. The second one is to modify Einstein's theory of gravitation. Therefore, cosmological models with many modified theories of gravity have always been close approximations of the real behavior of the universe. However, the rationale behind these modifications comes from Einstein-Hilbert action to obtain alternative theories of Einstein's theory of gravitation. Some relevant alternatives theories are Brans-Dicke(BD) theory, scalar-tensor theories of gravitation, $f(R)$ gravity \cite{carroll04, nojiri07, bertolami07}, $f(T)$ gravity  \cite{bengocheu09,linder10}, $f(G)$ gravity\cite{bamba10a, bamba10b, rodrigues14}, $f(R,G)$ gravity, where $R$, $T$ and $G$ are the scalar curvature, the torsion scalar and the Gauss-Bonnet scalar respectively. The recent generalisation of $f(R)$ gravity by introducing the trace of stress energy momentum tensor has become a most popular theory to represent the nature of expansion of the universe, known as $f(R,T)$ gravity proposed by Harko at al. in 2011 \cite{harko11}, where the matter Lagrangian consists of an arbitrary function of the curvature scalar $R$ and the trace of the energy-momentum tensor $T$. \\
In the wake of late time acceleration and existence of dark energy, various cosmological models have been formulated using modified theories of gravity. For example, Kaluza-Klein dark energy model and Bianchi Type VIh perfect fluid cosmological model in the $f(R, T)$ gravity have been investigated by Sahoo and Mishra \cite{Mishra14, Sahoo14}. The hydrostatic equilibrium configuration of neutron stars and strange stars were investigated in $f(R,T)$ gravity \cite{Moraes16}. The dynamics of magnetized string cosmological model have been discussed by Shri Ram and Chandel \cite{Ram15}. Sahoo et al. studied axially symmetric cosmological model in $f(R, T)$ gravity \cite{Sahoo014}. Moreover, the irregularity factor of self gravitating star due to imperfect fluid have been studied in $f(R, T)$ gravity by  Yousaf et al \cite{Yousaf16}.\\
 In addition, we have considered the bulk viscous fluid as our matter content. In literature, it plays an important role in the early evolution of the universe, for example \cite{Kumar15, Bali07}. In modern cosmology, the bulk viscosity is behaving more precisely, especially at inflationary phase for getting the accelerated expansion of the current universe\cite{Satish16}. It has been observed that viscosity can cause the qualitative behavior of solutions near the singularity without removing the total initial big bang singularity \cite{Belinskii76,murphy73}.    \\
In this paper, an exact solution of Einstein's field equations has been derived through a recent generalization of Friedmann-Lemaitre-Robertson-Walker (FLRW) metric space-time, called Bianchi type I  space-time, which is the simplest spatially homogeneous and anisotropic flat universe whose spatial sections are flat but the expansion or contraction rate is directional dependent. Also, the Bianchi type-I universe converges to Kanser universe near the singularity. Anisotropy of cosmic expansion has become a major concern for recent research community. This is because of the experimental data and its existence in the early phase which later approached to isotropic one with the passage of time along the age of the universe. Thus, it would be worthwhile to use Bianchi type-I universe in cosmological models in the account of anisotropic background. In literature, Bianchi type-I space-time has been used in various cosmological models for different aspects \cite{Sharif14, Zubair16, Rao13, Shamir15, Singh15}. \\
This work is organized in the following manner; In the second section, the metric and field equations are described and the solutions of the field equation with the physical and geometric behavior of the models are discussed. Finally, the third section ends with conclusions of this work.
 
\section{Field Equations and Solutions}
By considering the metric dependent Lagrangian density $L_m$, the respective field equation for $f(R,T)$ gravity is formulated from the Hilbert-Einstein variational principle in the following manner:
\begin{equation}
S=\int \sqrt{-g}\biggl(\frac{1}{16\pi G}f(R,T)+L_{m}\biggr)d^{4}x
\end{equation}%
where, $L_{m}$ is the usual matter Lagrangian density of matter source, $f(R,T)$ is an arbitrary function of
Ricci scalar $R$ and the trace $T$ of the energy-momentum tensor $T_{ij}$ of the matter source and $g$ is the determinant of the metric tensor $g_{ij}$. The energy-momentum tensor $T_{ij}$ from Lagrangian matter is defined in the form
\begin{equation}
T_{ij}=-\frac{2}{\sqrt{-g}}\frac{\delta (\sqrt{-g}L_{m})}{\delta g^{ij}}
\end{equation}%
and its trace is $T=g^{ij}T_{ij}$.\newline
Here, we have assumed that the matter Lagrangian $L_{m}$ depends only on the
metric tensor component $g_{ij}$ rather than its derivatives. Hence, we
obtain
\begin{equation}
T_{ij}=g_{ij}L_{m}-\frac{\partial L_{m}}{\partial g^{ij}}
\end{equation}%
By varying the action $S$ in Eq. (1) with respect to $g_{ij}$, the $f(R,T)$
gravity field equations are obtained as
\begin{equation}
F(R,T)R_{ij}-\frac{1}{2}f(R,T)g_{ij}+(g_{ij}\Box -\nabla _{i}\nabla
_{j})F(R,T)\\= 8\pi T_{ij}-\mathcal{F}(R,T)T_{ij}- \mathcal{F}(R,T)\Theta _{ij}
\end{equation}%
where,
\begin{equation}
\Theta _{ij}=-2T_{ij}+g_{ij}L_{m}-2g^{lm}\frac{\partial ^{2}L_{m}}{\partial
g^{ij}\partial g^{lm}}
\end{equation}
Here, $F(R,T)=\frac{\partial f(R,T)}{\partial R}$, $\mathcal{F}(R,T)=\frac{%
\partial f(R,T)}{\partial T}$, 
$\Box \equiv \nabla ^{i}\nabla _{i}$ where $%
\nabla _{i}$ is the co-variant derivative.\newline
Contracting Eq. (4), we get
\begin{equation}
F(R,T)R+3\Box F(R,T)-2f(R,T)= 
(8\pi -\mathcal{F}(R,T))T-\mathcal{F}(R,T)\Theta
\end{equation}%
where $\Theta =g^{ij}\Theta _{ij}$.\newline
From Eqs (4) and (6), the $f(R,T)$ gravity field equations takes the form
\begin{equation}
F(R,T)\biggl(R_{ij}-\frac{1}{3}Rg_{ij}\biggr)+\frac{1}{6}f(R,T)g_{ij}=
8\pi -\mathcal{F}(R,T) \biggl(T_{ij}-\frac{1}{3}Tg_{ij}\biggr)-\mathcal{F}(R,T)\biggl(%
\Theta _{ij}-\frac{1}{3}\Theta g_{ij}\biggr)+\nabla _{i}\nabla _{j}F(R,T)
\end{equation}%
It is important to note that the physical nature of the matter field is very important for the field equations of $f(R,T)$ gravity through the tensor $\theta_{ij}$. So for different choices of matter, one can construct several  cosmological models addressing with a different explicit form of $f(R,T)$ such as:
\begin{itemize}
\item $f(R,T)=  R+2f(T)$ 
\item $f(R,T)= f_{1}(R)+f_{2}(T)$
\item $f(R,T)= f_{1}(R)+f_{2}(R)f_{3}(T)$
\end{itemize}
Various cosmological models have been constructed in different choices of $f(R,T)$ gravity in several aspects \cite{mahanta14, singh14, sahoo15, sahoo17, sahoo/2017}.
Here, we consider the spatially homogeneous LRS Bianchi type-I metric as
\begin{equation} \label{1}
ds^{2}=dt^{2}-a_1^{2}dx^{2}-a_2^2(dy^{2}+dz^{2})
\end{equation}
where $a_1, a_2$ are functions of cosmic time $t$ only.\\
The energy momentum tensor in matter of bulk viscus fluid is taken as in this form
\begin{equation}\label{2}
T_{ij}=(\rho+\overline{p})u_{i}u_{j}-\overline{p}g_{ij}
\end{equation}
where $u^i=(0,0,0,1)$ is the four velocity vector in co-moving coordinate system satisfying $u_iu_j=1,$
\begin{equation}\label{3}
\overline{p}=p-3\xi H
\end{equation}
is the bulk viscous pressure which satisfies the linear equation of state $p=\gamma \rho, 0\leq \gamma \leq 1$, $\xi$ is the bulk viscous coefficient, $H$ is Hubble's parameter, $p$ is pressure and $\rho$ is the energy density.\\
 The trace of energy momentum tensor is given as
\begin{equation}\label{4}
T=\rho-3\overline{p}
\end{equation}
\hspace*{-2mm}The field equations of $f(R,T)$ gravity using linear case $f(R,T)=R+2f(T)$ can be written as
\begin{equation} \label{5}
R_{ij}-\frac{1}{2}Rg_{ij}=8\pi T_{ij}-2(T_{ij}+\Theta_{ij})f'(T)+f(T)g_{ij}
\end{equation}
\hspace*{1mm}where $f(T)=\alpha T$, $\alpha$ is an arbitrary constant.\newline

\hspace*{-2mm}The field equations (\ref{5}) for the metric (\ref{1}) are obtained as

\begin{eqnarray}
-2\frac{\ddot{a_2}}{a_2}-\frac{\dot{a_2}^{2}}{a_2^{2}}=(8\pi+3\alpha)\overline{p}-\alpha \rho  \label{6}\\
-\frac{\ddot{a_1}}{a_1}-\frac{\ddot{a_2}}{a_2}-\frac{\dot{a_1}\dot{a_2}}{a_1a_2}=(8\pi+3\alpha)\overline{p}-\alpha \rho \label{7}\\
2\frac{\dot{a_1}\dot{a_2}}{a_1a_2}+\frac{\dot{a_2}^{2}}{a_2^{2}}=(8\pi+3\alpha) \rho-\alpha \overline{p} \label{8}
\end{eqnarray}
\hspace{2mm}where dots represent the derivatives with respect to time $t.$\\
\hspace*{2mm}The deceleration parameter is defined as
\begin{equation}\label{9}
q=-\frac{a\ddot{a}}{\dot{a}^2}
\end{equation}
\hspace*{2mm}where $a$ is the average scale factor.\\
\hspace*{2mm}Integrating the above equation we get
\begin{equation}\label{10}
a(t)=e^\delta exp \bigg[\int \frac{dt}{\int(1+q)dt+\eta}\bigg]
\end{equation}
\hspace*{2mm}where $\delta$ and $\eta$ are integration constants.\\
\hspace*{2mm}Here, we have three field equations involving four
unknown  parameters as $a_1, a_2, \overline{p} \ \&\ \rho$. In order
to solve these undetermined system of equations, we assume
\hspace*{2mm}a time dependent deceleration parameter $q$ proposed by Abdussattar and Prajapati \cite{Abdussatter11} given as
\begin{equation}\label{11}
q=-\frac{k_1}{t^2}+(k_2-1)
\end{equation}
where $k_1>0$, $k_2>1$ are constants.\\
As per the observational results \cite{Riess98, Bennet2003}  the universe exhibits phase transition, that means the transition occurs from the past decelerating phase to the recent accelerating expansion phase. The physics of this phenomena can be conducted through a geometrical parameter called as deceleration parameter ($q$) which describes the acceleration or deceleration behavior of the universe depending on the negative or positive value. Due to this reason, it has great impacts on the behavior of the cosmological model in the expansion of the universe. Hence our choice of variable $q$ is physically acceptable. \\
  From equation (\ref{11}) one can observe that the
deceleration parameter $q\rightarrow -\infty $ at $t=0$ and it reduces to zero at $t=\sqrt{\frac{k_1 }{k_2 -1}}$. The period of
acceleration depends on $k_1 $ and $k_2 $ which is an increasing function of time as per the accelerated phenomenology of the present universe. At large value of $t$, the model is decelerating for $q \rightarrow \beta-1$. Thus, there is a restriction on $\beta$ as $1\leq \beta \leq 2$. More particular, at $\beta=1$ the
universe has an accelerated expansion throughout the evolution for negative value of $q$. From Fig-1, it can be seen that the deceleration parameter is completely negative with large value of $k_1$ and it provides the
major evidence for our model is accelerating one.  \newline
\begin{figure}[h!]
\centering
\includegraphics[width=75mm]{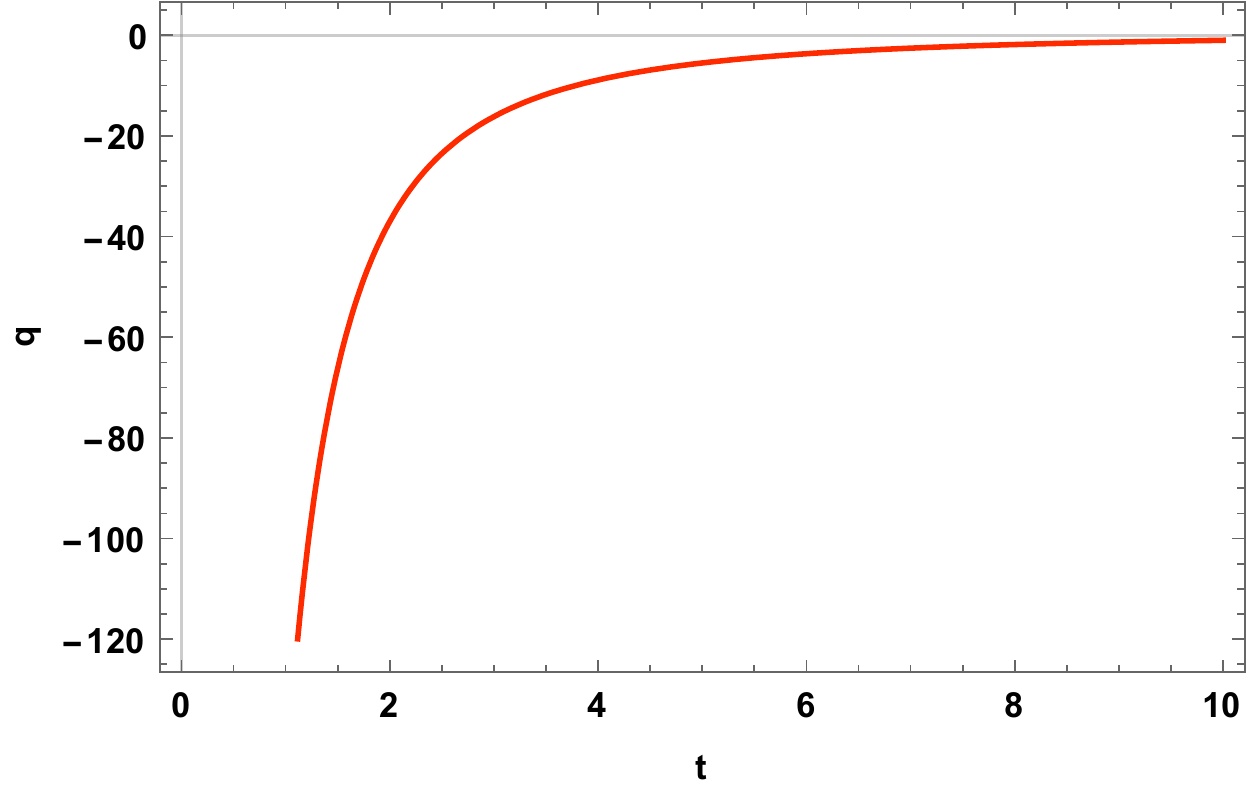}
\caption{$q$  vs. cosmic time $t$ with $k_1=150$, $k_2=1.5$}\label{fig1}
\end{figure}
 Using equation ($\ref{11}$) in equation ($\ref{10}$) and by considering  $\delta=\eta=0$ for simplicity. we get,
\begin{equation}\label{12}
a(t)=\bigg(t^2+\frac{k_1}{k_2}\bigg)^{\frac{1}{2k_2}}
\end{equation}

From field equation (\ref{5}) to (\ref{7}) we obtain
\begin{equation}\label{13}
\frac{a_1}{a_2}=c_2  exp\bigg(\int\frac{c_1}{a^3}dt \bigg )
\end{equation}
Here, the volume $V=a^3=a_1a_2^2$ and $ c_1$, $c_2$ are  integration constants.\\
To find the values of the metric potentials $a_1, a_2$ we consider a
particular case $k_2=\frac{3}{2}$ in equation ($\ref{12}$), so that
\begin{equation}\label{14}
a(t)=(t^2+b^2)^{\frac{1}{3}}
\end{equation}
where $b^2=\frac{2k_1}{3}.$
and the mean Hubble parameter is given as
\begin{equation}\label{15}
H=\frac{1}{3}(H_1+2H_2)=\frac{2t}{3}(t^2+b^2)^{-1}
\end{equation}
The use of this deceleration parameter in this model is more appropriate as it provides a continuous expansion. It can be observed through the behavior of scale parameter and mean Hubble parameter given in equation (\ref{14}) and (\ref{15}) as shown in fig-2 and fig-3 respectively. It can be seen that from fig-2, $a\rightarrow \infty$ at $t\rightarrow \infty$, while in fig-3, Hubble parameter is a positive decreasing function with time, at large value of $t$ it approaches to zero.\\
\begin{figure}[h!]
\begin{minipage}[t]{.45\textwidth}
\centering
      \includegraphics[width=75mm]{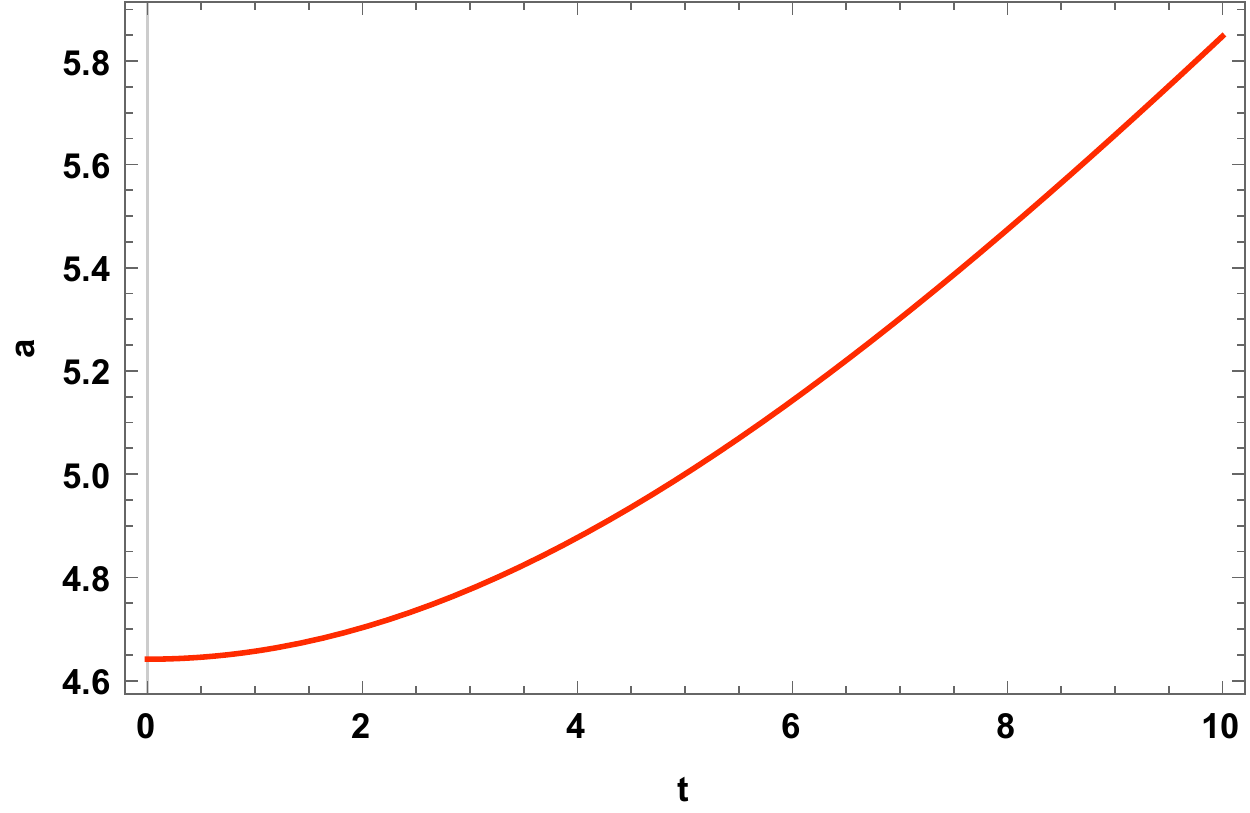}
\caption{$a$ vs. cosmic time $t$ with $k_1=150$ and $k_2=1.5$.}\label{fig2} 
  \end{minipage}
\begin{minipage}[t]{.45\textwidth}
\centering
\includegraphics[width=75mm]{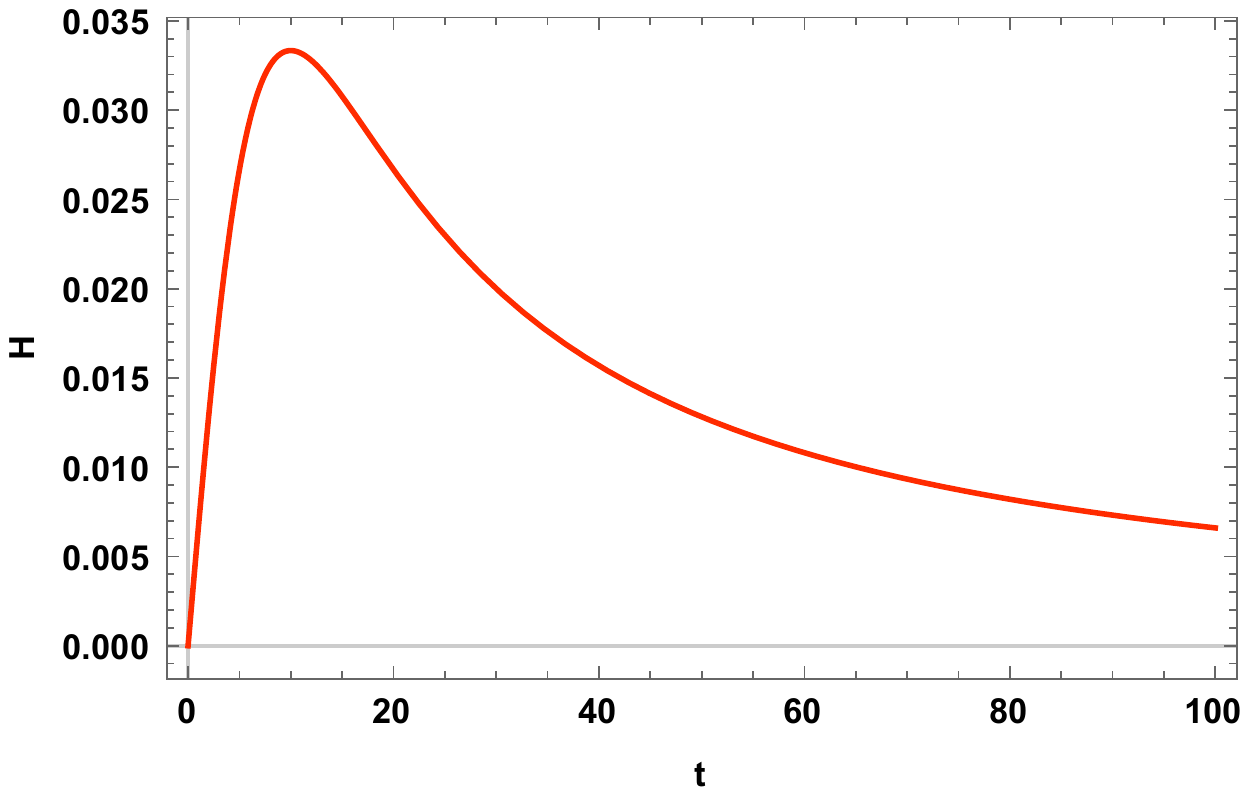}
\caption{$H$  vs. cosmic time $t$ with $k_1=150$, $k_2=1.5$}\label{fig3}
  \end{minipage}
\end{figure}

If we consider $k_1=0$ in equation ($\ref{12}$) then
\begin{equation}\label{16}
a(t)=t^{\frac{1}{k_2}}
\end{equation}
which yields a constant deceleration parameter $q=k_2-1$ throughout the
universe \cite{Berm83}.\\  
  From equation ($\ref{13}$), we obtain the metric  potentials as:
\begin{equation}\label{17}
a_1=c_2^{2/3} \sqrt[3]{b^2+t^2} e^{\frac{2 c_1 \tan ^{-1}\left(\frac{t}{b}\right)}{3 b}}
\end{equation}
\begin{equation}\label{18}
a_2=\frac{\sqrt[3]{b^2+t^2} e^{-\frac{c_1 \tan ^{-1}\left(\frac{t}{b}\right)}{3 b}}}{\sqrt[3]{c_2}}
\end{equation}
By solving the field equations ($\ref{7}$) and ($\ref{8}$), the
values of $\rho$ and $\overline{p}$ are obtained as:
\begin{equation}\label{19}
\rho=\frac{8 \pi  t^2-\alpha  \left(b^2-3 t^2\right)-(\alpha +2 \pi ) c_1^2}{6 \left(\alpha ^2+6 \pi  \alpha +8 \pi ^2\right) \left(b^2+t^2\right)^2}
\end{equation}
\begin{equation}\label{20}
\overline{p}= -\frac{(3\alpha +8\pi)b^2+(\alpha+2\pi)c_1^2-\alpha t^2}{6 \left(\alpha ^2+6 \pi  \alpha +8 \pi ^2\right) \left(b^2+t^2\right)^2}
\end{equation}
The values of coefficient of bulk viscosity $\xi$ and the Equation of State(EOS) parameter are
\begin{equation}\label{21}
\xi=\frac{b^2(8\pi-\alpha(\gamma-3))-(\alpha+2\pi)(\gamma-1)c_1^2}{12\left(\alpha ^2+6 \pi  \alpha +8 \pi ^2\right)t \left(b^2+t^2\right)}+
\frac{t^2(3\alpha \gamma-\alpha+8\pi \gamma)}{12\left(\alpha ^2+6 \pi  \alpha +8 \pi ^2\right)t \left(b^2+t^2\right)}
\end{equation}
\begin{equation}\label{22}
\omega=\frac{\rho}{\overline{p}}=-\frac{8 \pi  t^2-\alpha  \left(b^2-3 t^2\right)-(\alpha +2 \pi ) c_1^2}{(3\alpha +8\pi)b^2+(\alpha+2\pi)c_1^2-\alpha t^2}
\end{equation}
From equation \ref{18} to \ref{21}, we have observed that the energy density is positive throughout the universe and decreases with time and at the late time it approaches to zero i.e $\rho\rightarrow 0$ as $t\rightarrow \infty$. The behavior of bulk viscous pressure ($\overline{p}$)  against cosmic time t revealed that bulk viscous pressure is a negative increasing function starting off from a large negative value and it is tending to zero with the evolution of time i.e. $\overline{p}\rightarrow 0$ at $t\rightarrow\infty$. Recent observational astronomy has predicted a specified range for the value of equation of state (EOS) parameter ($\omega$), which is a function of pressure and energy density i.e. $\omega=\frac{\rho}{\overline{p}}$, as $-1\leq \omega \leq 0$, in which case the evolution of the universe is under acceleration. In the simplest case, the cosmological constant($\Lambda$) appears for a particular value of EOS parameter; $\omega=-1$. Similarly, phantom model and quintessence model also arise in cosmology when $ \omega\leq -1$ and $\omega\geq-1$ respectively. Here we have covered a quintessence cosmological model, where the accelerated expansion of the universe was indeed observed. 
Some other physical parameters of this model are obtained as\\

Spatial volume(V):
\begin{equation}\label{23}
V=a_1a_2^2=(t^2+b^2)
\end{equation}
Expansion scalar:
\begin{equation}\label{24}
\theta=3H=(2t)(t^2+b^2)^{-1}
\end{equation}
Shear scalar:
\begin{equation}\label{25}
\sigma^2=\frac{c_1^2}{3 \left(b^2+t^2\right)^2}
\end{equation}
Mean anisotropic parameter:
\begin{equation}\label{26}
\Delta=6 \frac{\sigma^2}{\theta^2}= \frac{c_1^2}{2 t^2}
\end{equation}
In this model, the spatial volume is zero at initial time $t=0$ but it will increase along with time. The expansion scalar is a decreasing function of time. It starts from the infinite value at the initial epoch of time and then approaches to zero at a later stage. $\Delta$ is the mean anisotropy parameter of this model given in equation (\ref{26}). We observe that it is a decreasing function of time i.e at late time when $t\rightarrow\infty, \Delta\rightarrow 0$. The dynamics of the mean anisotropic parameter in our model shows a transition from initial anisotropy to isotropy in the current epoch which is in good agreement with recent observations.
\section{Conclusion}
In order to obtain a physically viable cosmological model, it is necessary that it can reproduce the several different epochs in the present evolution of the universe. From the viewpoint of the fluid description, and the current cosmological observational data, the energy
components of the universe are treated as imperfect fluids due to the presence of a bulk viscosity in them. Many works 
are described with viscous fluid to study the evolution of the universe as per the section-I review. As per the literature, the cosmic bulk viscosity is treated as a viable candidate for providing a theoretical explanation for early and late time expansion of the universe. Therefore, in this work, we have investigated an accelerated cosmological model in an anisotropic universe using the linear frame of $f(R,T)$ gravity theory. A new class of Bianchi-I bulk viscous model has been observed, where the time varying deceleration parameter plays an important role to get exact solutions of field equations. Our model also represents expanding and shearing which approaches to isotropy for large values of $t$. This is consistent with the behavior of the present universe. Through our work, we hope to present a better understanding of the evolution of the universe in the Bianchi type-I space time within the framework of $f(R,T)$ gravity which may be useful to study the role of bulk viscosity in the expansion and evolution of the universe.
 
\acknowledgments 
The authors (PS and VRR) would like to thank their academic supervisor Prof.P. K. Sahoo for  helpful discussions and constant encouragement for research. Also, the authors  would like to acknowledge DST, New Delhi, India for providing facilities through DST-FIST lab, Department of Mathematics, where a part of this work was done.\\
 The authors are very indebted to the editor and the anonymous referees for illuminating suggestions that have significantly improved our paper in terms of research quality as well as the presentation.

\end{document}